# Prime Factorization Using Magnonic Holographic Device

Y. Khivintsev[1], M. Ranjbar[2], D. Gutierrez[2], H. Chiang [2], A. Kozhevnikov[1], Y. Filimonov[1], and A. Khitun[2]

[1]*Kotel'nikov Institute of Radioengineering and Electronics of Russian Academy of Sciences, Saratov Branch, Saratov, Russia, 410019*

[2]*Electrical Engineering Department, University of California - Riverside, Riverside, CA, USA, 92521*

**Abstract**

Determining the prime factors of a given number *N* is a problem, which requires super-polynomial time for conventional digital computers. A polynomial-time algorithm was invented by P. Shor for quantum computers. However, the realization of quantum computers is associated with significant technological challenges in achieving and preserving quantum entanglement. Prime factorization can be also solved by using classical wave interference without quantum entanglement. In this work, we present experimental data showing prime factorization by utilizing spin wave interference. The prime factorization includes three major steps. First, general type computer calculates the sequence of numbers $m^k \mod(N)$, where *N* is the number to be factorized, m is a randomly chosen positive integer, and k=0,1,2,3,4,5,6 .. . Next, the period of calculated sequence *r* is determined by exploiting spin wave interference. Finally, the general type computer finds the primes based on the obtained *r*. The experiment on the period finding was accomplished on the 6-terminal $Y_3Fe_2(FeO_4)_3$ device. We chose number 15 for a test and found its primes in a sequence of measurements. The obtained data demonstrate an example of solving a prime factorization problem using classical wave interference. We discuss the physical and technological limitations of this approach, which define the maximum size of *N* and the computational speed. Though this classical approach cannot compete with the quantum algorithm in efficiency, magnonic holographic devices can be potentially utilized as complementary logic units aimed to speed up prime factorization with classical computers.

**One Sentence Summary:** Our results show an example of solving the prime factorization problem by using spin wave interference. Potentially, magnonic holographic devices may complement digital logic circuits in special task data processing.

I. Introduction

Prime factorization is the process of finding the set of prime numbers which multiply together to give the original integer *N*. The most naive approach to the this problem is a step by step checking all possible numbers from 2 to $\sqrt{N}$, where the number of operations increases exponentially with the increase of the input state *N*. Even implementation of the most efficient algorithms (e.g. Fermat's factoring algorithm (*1*)) would take an enormous amount of operations in a supercomputer to find the primes of a large *N*. This fact explains the use of prime encoding in information security (*2*).

Quantum computing has emerged as a promising computational paradigm in part due to its ability to solve prime factorization problem more efficiently than with conventional digital computers. P. Shor has developed a polynomial-time quantum algorithm for the factoring problem, and its computational complexity has been proved to be $O((\log N)^2 (\log \log N)(\log \log \log N))$, which provides a fundamental advantage over any type of digital type computing (*3*). Though the fundamental advantage of quantum computing is undisputable, its practical realization is associated with multiple technological challenges required for quantum entanglement implementation.

The prime factorization problem can be also solved by using classical wave interference. This approach has been intensively studied in optics (*4*). In general, finding the factors of a given number N can be accomplished via the use of the truncated sum:

$$\Pi_N^{(M)}(s) = \frac{1}{M}\sum_{m=1}^{M}\exp\left(-2\pi i m^k \frac{N}{s}\right), \qquad (1)$$

where k is an integer and M is the number of terms in the sum. The argument *s* scans through all integers between 1 and $\sqrt{N}$ for possible factors. The capability of the sum of Eq. (1) to factor numbers originate from the fact that for an integer factor q of *N* with $N=q\times s$, all phases in Π are integer multiples of 2π. Consequently, the terms add up constructively and yield Π=1. When *s* is not a factor, the phases oscillate rapidly with m, and Π takes on small values. In this interference pattern, larger truncation parameter M leads to better convergency. In principle, the first several terms of the sum are sufficient to discriminate factors from non-factors. There are several specific procedures utilizing time delay, pulse train, and frequency beats allowing to do factorization with classical waves (*4*). In general, any type of classical waves can be exploited, where each of the possible approaches possesses its own advantages and shortcomings.

Recently, it was discussed the possibility of using Magnonic Holographic Memory (MHM) devices for prime factorization (*5*). MHM is a type of holographic memory which utilizes spin wave interference. The primary projected application of MHM is parallel read-out of magnetic bits (*6*). MHM also possesses great potential for data processing (e.g. pattern recognition) (*7*). The advantages of using spin waves include scalability and compatibility with conventional electronic devices. In this work, we present experimental data on finding the primes of the number 15 by using a 6-terminal MHM. The rest of the work is organized as follows. In the next Section II, we briefly describe the prime factorization procedure using MHM. In Section III, we describe experimental setup and present experimental data. Physical limitations of using spin wave devices are discussed in Section IV.

## II. Prime Factorization with MHM

The whole procedure includes three major steps. First, the general type computer calculates the sequence of numbers $m^k \mod(N)$, where m is a randomly chosen positive integer, k = 0,1,2,3,4,5,6... Secondly, the obtained numbers are converted into the spin waves in MHM. The waves are excited sequentially, where each new number in the modular sequence adds a spin wave to the interference pattern. The output detects the phase and the amplitude of the signal produced by the spin wave interference. In this scenario, the output of MHM oscillates with the period $r$, where $r$ the period of the calculated sequence of numbers. Finally, the primes are found by the general type computer based on the obtained period $r$: $\gcd(m^{r/2} + 1, N)$ and $\gcd(m^{r/2} - 1, N)$. This procedure is illustrated in Fig.1.

For example, the procedure of finding the primes of the number 15 is the following. We choose m=7 and calculate $7^k \bmod(15)$ for k=0,1,2,3,..,20. The calculations of the mod function are done by the general type computer. The first calculated number is 7. It is converted into a spin wave with amplitude $A_1 = A_0$ and phase $\phi_1$. The second calculated number is 4. It is a new number in the sequence, which is converted into a spin wave with amplitude $A_2 = A_0$ and phase $\phi_2 \neq \phi_1$. As the next two numbers (i.e. 13 and 1) are calculated, two more waves are added to the input. As in the previous cases, these waves have the same amplitude $A_0$, and phases $\phi_3$ and $\phi_4$, respectively. Thus, each distinct number in the modular function is converted into a spin wave with a distinct phase (i.e. $\phi_1 \neq \phi_2 \neq \phi_3 \neq \phi_4$). The fifth computation gives number 7, which has been already related to the spin wave of phase $\phi_1$. In this case, the amplitude of the spin wave at port 1 is doubled $A_1 = \eta A_0$, where $\eta$ is some positive number. And so on, every new number in the sequence adds a new spin wave signal, while the amplitude of the signal is multiplied by $\eta$ every time the same number is found. The phase of the output oscillates with the same period as the period of the modular function. For instance, the phase of the output is the same for step 4 (all antennas generate spin waves of amplitude $A_0$), step 8 (all antennas generate spin waves of amplitude $\eta A_0$), step 12, and so on with period $r = 4$. That is the key idea of the MHM proposed approach for finding the period of the modular function by exploiting spin wave interference. Finally, the primes are found by the general type computer based on the obtained period $r$: $\gcd(m^{r/2} + 1, N)$ and $\gcd(m^{r/2} - 1, N)$.

**III. Experimental data`**

The photo of MHM device and the schematics of the experimental setup are shown in Fig.2. Fig.2(a) shows the device packaged. The device consists of a double-cross structure (as shown in Fig.2(b)) made of yttrium iron garnet $Y_3Fe_2(FeO_4)_3$ (YIG) epitaxially grown on gadolinium gallium garnet $Gd_3Ga_5O_{12}$ substrate with crystallographic orientation (111). This material is chosen due to its long spin wave coherence length and relatively low damping(*8*), which makes it the best candidate for room temperature spin wave device prototyping. YIG film has ferromagnetic resonance (FMR) linewidth $2\Delta H \approx 0.5$ Oe, saturation magnetization $4\pi M_s = 1750$ G, and thickness= 3.6μm. The length of the whole structure is 3 mm, the width of the arm is 360 μm. There are six micro-antennas fabricated on the top of the YIG waveguides. These antennas are aimed to excite spin waves and to detect the inductive voltage produced by the interfering spin waves. The excitation of the spin waves are accomplished by passing AC electric current through the antenna contour, which, in turn, generate AC magnetic field and may excite propagating spin waves in the waveguide under the resonance conditions. The inductive voltage is a result of the spin wave interference, where the propagating spin waves alter the magnetic flux from the surface. The inductive voltage has maxima when spin waves are coming in phase and show minima when spin waves are coming out-of-phase. The set of equations connecting the amplitudes and the phases of multiple spin waves and the resultant inductive voltage can be found in Ref. (*9*). Both the phase and the amplitude of the inductive voltage are measured at the output. The details of the inductive voltage measurements using micro-antennas can be found in Ref. (*10, 11*).

In our experiments, we used just one antenna for the output detection (i.e. antenna #1 in Fig.2(c)), and four antennas for spin wave excitation (i.e. antennas numbered 2,3,4,5 in Fig.2(c)). The input and the output micro-antennas are connected to a Keysight Technologies programmable network analyzer (PNA) N5221A-217. The PNA generates an input RF signal and measures the $S_{12}$ parameters showing the amplitude and the phase of the transmitted signal. The input from PNA is split between the four inputs via the system of splitters, phase shifters and attenuators as shown in Fig. 2(c). The output voltage from antenna #1 is amplified +15 dBm. The YIG device is placed inside an electro-magnet allowing variation in the bias magnetic field from -1000 Oe to +1000 Oe. The external magnetic field is directed in-plane as shown in Figure 1(B). Before the experiment, it was found the region in the frequency-bias magnetic field space where both backward volume magnetostatic spin waves (BVMSW) and magnetostatic surface spin waves (MSSW) can propagate. The latter is critically important for the operation of the cross-shape magnonic devices as the input spin waves initially propagate perpendicular to the bias field (MSSW), while reach the output propagating along the external magnetic field (BVMSW). The most prominent overlap takes place at the frequency 4.165 GHz and 760 Oe bias magnetic field. All experimental data were collected at this particular combination.

Following the procedure described in Section II, we sequentially excited spin waves by the four antennas and detected the phase and the amplitude of the output inductive voltage at port 1. Prior to the experiment, we used the system of the attenuators as depicted in Fig.2(b) to equalize the output inductive voltage produced by the independently working antennas (i.e. ports #2,3,4 and 5). We setup the following phase difference between the input ports: 0π (port #2), π/3 (port #3), 2π/3 (port #4), and π/2 (port #5). Then, we carried out 20 sequential measurements. The collected data are summarized in Table I. It shows the number of spin wave generating antennas, the amplitude and the phase of the produced inductive voltage. During this experiment, we changed the amplitude of the spin wave generating antenna. For instance, the amplitude of the spin wave signal at port # 2 was reduced approximately two times ($\eta$=0.5) at the measurement number 5. In Table II, we show experimental data on the amplitude and the phase of the inductive voltage produced by the independently working antennas during the experiment. In Fig.3, we plotted the phase of the output for the 20 sequential measurements (red markers and red curve). As expected, the phase of the output voltage oscillates with a period $r$ = 4, which is the period of the calculated modular function 7,4,13,1,7,4,13,1…

The period of the modular function can be found via the amplitude measurements as well. In Fig.4(a), we plotted the power of the output signal for the 20 measurements. The blue curve and blue markers depict the output power $P_{int}$ for the interfering spin waves (i.e. from Table I). The green curve and green markers depict the sum power $P_{sum}$ as would be produced by the independently working antennas (i.e. data from Table II). In both cases, there is an increase of the output amplitude as we increases the number of spin wave generating antennas. The output amplitude decreases as we start to reduce input power (i.e. measurement #5). In Fig. 4(b), we plotted the normalized output $P_{int}/P_{sum}$, which shows oscillation with a period $r = 4$. Thus, both the phase and the amplitude measurements provide the same result $r = 4$, which allows us to find the primes of number 15: 5 and 3 (i.e. gcd($7^{4/2} + 1$, $15$) and gcd($7^{4/2} - 1$, $15$)).

We want to note the variations of the phase and the amplitude of the input signals during the experiment. The phase difference between the spin wave generating ports deviated about 2%, while the amplitude variation exceeded 10%. In Fig.3 and Fig.4(b), we plotted the results of numerical modeling (i.e. the black curve and the black markers) showing the output signal in the ideal case of zero phase/amplitude variations. There is a good agreement between the experimental and theoretical data is observed in the case of phase measurements as shown in Fig.3. There is more discrepancy observed in the case of amplitude measurements as shown in Fig.4(b). However, in both cases, we can find the period of the output oscillation despite of significant input signal variations.

**IV. Discussion**

The experimental data presented above demonstrate an example of period fining by utilizing spin wave interference. The same procedure can be accomplished with any other type of classical waves (e.g. optical, acoustic, gravitational, etc.). In this Section, we would like to discuss the advantages and shortcomings of the spin wave approach. Operation wavelength, group velocity, coherence length and thermal noise level are the key physical parameter defining the scalability, speed, and the maximum size of number *N*, which can be factorized by classical wave interference. Scalability is one of the main appealing properties of MHM. Spin wave wavelength can be reduced to the nanometer range, which translates in the possibility of making scalable MHM devices designed for period finding. For instance, the minimum are of the device shown in Fig.2 (b) is 5λ×3λ, where λ is the operational wavelength. Taking λ=100nm, the area of the MHM devices is $1.5 \times 10^{-9}$ cm$^2$, which is less than the size of a 32 bit adder built of 14nm CMOS (*9*). It is also important to estimate how much time it would take for the whole period finding procedure. It has several components including the time required for classical computer to calculate the $m^k$mod(N), the time required for spin wave generation and detection. The time required for spin wave signal generation is limited by the RC delay (R is the resistance and C is the capacitance of the circuit) of the conducting contours. The propagation time of the spin wave signal is defined by the length of the device and the spin wave group velocity. In our case, the length of the device is 3mm, and group velocity of the magnetostatic spin waves in YIG is about $3\times10^4$ m/s. So the time of the signal propagation is 0.1μs. Then, it will take at least one period of oscillation (0.1ns) to detect the amplitude and the phase of the output signal. The longest among the all above mentioned is the time required for signal to propagate from the input antenna to the output detector.

The main question is related to the maximum size of N, which can be factorized using spin waves. In general, one period of the function $m^k$mod(N) may contain *n* district numbers, where the frequency of the same number appearing $f_n$. On the first look, the finding of a period with large *n* will require *n* spin wave generating antennas, where $\phi_n$ is assigned to the each number in the sequence. However, this problem can be resolved by utilizing one of the antennas as a "memory unit". For instance, the output at measurement #4 has amplitude 1.88mW and the phase 42 degrees. The memory port can be adjusted to provide the same output, while the other antennas are used to continue the procedure. In this scenario, the 6-port MHM as shown in Fig.2(a) can be used for finding the period with *n* » 6. It is also important to take into consideration the optimum way of input phase $\phi_n$ choosing: $\phi_{n+1} = 2\pi - \phi_n$. The latter will prevent the output amplitude from growing (i.e. the first 4 measurements in our case). In any case, the maximum period $r_{max}$ is defined by the accuracy which we can measure the amplitude and the phase of the output as follows:

$$r_{max} = n \times f_n = \frac{2\pi}{\Delta\phi} \times \frac{A_{max}}{\Delta A}, \qquad (2)$$

where $\Delta\phi$ is the minimum detectable phase change, $A_{max}$ is the maximum amplitude of the spin wave signal, which can be generated or sustained by the device, $\Delta A$ is the minimum change of the output amplitude, which can be detected. The first term in the Eq.(2) corresponds to the number of different terms in the modular sequence (i.e. each distinct number correspond to a certain input phase), while the second term in Eq.(2) corresponds to the number of times this number appears within a period. In the ultimate limit, $r_{max}$ is limited by the thermal noise. The flicker *1/f* noise level in ferrite structures usually does not exceed -130 dBm (*12*). In this work, the maximum power of the spin wave signal was about 2.0 mW (3 dBm). Thus, the maximum frequency of repetition of any distinct number (i.e. $A_{max}/\Delta A$) is $10^{11}$. The phase control accuracy of 2% allows us to use 180 distinct phases φ for assigning to the different numbers in the sequence. In this ultimate case, the maximum period which can be found by using spin wave interference can be estimated as $r_{max}=1.8\times10^{13}$. Though this number may look impressive, the minimum time required for $10^{13}$ measurements ($10^{13}\times10^{-7}$s=$10^{5}$ s) exceeds two hours without taking into consideration the time required for the general type computer for modular function calculation. It should be clear that the classical wave-based logic circuits cannot compete with quantum computers in efficiency. However, the use of these devices may complement conventional digital logic circuits in special type data processing (i.e. prime factorization).

**V. Conclusions**

We have presented experimental data showing an example of period finding by utilizing spin wave interference. The experiment was accomplished on the 6-terminal YIG device where spin waves were excited and detected by the set of micro antennas. Four antennas were used to excite continuous spin wave signals where the phase of the signal correspond to the distinct numbers and the amplitude relates to the frequency of the number appearing in the sequence. The output was detected via the inductive voltage measurements. As an example, we found the period of the function $7^k\mod(15)$, where k=1,2,3,… 20. The obtained experimental data show a good agreement with the results of numerical modeling despite the significant variation of the input phases and amplitudes. The maximum period, which can be found using spin wave device, is limited by the *1/f* noise level, while the speed of a single computation is limited by the spin wave group velocity. In the ultimate limit, spin wave devices can be used for finding relatively large periods up to $10^{13}$ due to the low noise level in non-conducting ferrite materials. The other advantages of using spin waves include scalability and compatibility with conventional electronic logic circuits. Though classical wave-based device cannot compete with quantum computers in efficiency, they may be used as a complementary logic circuits aimed to speed up the prime factorization procedure with conventional digital computers.

**Materials and Methods**

$Y_3Fe_2(FeO_4)_3$ (YIG) is the most convenient material for magnonic holographic devices prototyping as it possesses relatively long (~1cm) spin wave coherence length at room temperature. The excitation and detection of the spin waves is accomplished by the set of micro antenna fabricated directly on the top of the YIG waveguides. The antennas can be fabricated by the conventional optical lithography. The operational frequency and the bias magnetic field should be adjusted to the region where both backward volume magnetostatic spin waves (BVMSW) and magnetostatic surface spin waves (MSSW) can propagate. The input and the output micro-antennas are connected to a vector network analyzer to measure the $S_{12}$ parameters showing the amplitude and the phase of the transmitted signal. There no special programs or algorithms required for the mathematical analysis of the obtained data.

**Acknowledgment**

This work was supported in part by the FAME Center, one of six centers of STARnet, a Semiconductor Research Corporation program sponsored by MARCO and DARPA.

**Figure Captions:**

Figure 1. Illustration of the prime factorization procedure. The general type computer calculates the sequence of numbers $m^k \mod(N)$, where m is a randomly chosen positive integer, k(0,1,2,3,4,5,6 ..). Next, the obtained numbers are converted into the spin waves in the MHM device. The waves are excited sequentially, where each new number in the modular sequence adds a spin wave to the interference pattern. The output of MHM oscillates with the period *r*, where *r* the period of the calculated sequence of numbers. Finally, the primes are found by the general type computer based on the obtained period *r*: $gcd(m^{r/2} + 1, N)$ and $gcd(m^{r/2} - 1, N)$.

Figure 2. (a) Photo of the packaged device. (b) Photo of the double-cross $Y_3Fe_2(FeO_4)_3$ structure. The length of the structure is 3mm, the width of the arm in 360µm, the thickness of the YIG film is 3.6µm. (c) The schematics of the experimental setup. There are five micro-antennas fabricated on top of the YIG structure. The antennas are connected to the programmable network analyzer (PNA) via the set of splitters (depicted as S). Four antennas (#2,3,4,5) are to excite spin waves, and one antenna (#1) is to detect the inductive voltage produced by the spin wave interference. There is a system of phase shifters (depicted as P) and attenuators (depicted as A) to control the phase and the amplitude of the spin wave signals.(d) Schematics of the YIG structure with 5 micro-antennas. The device is placed inside an electro-magnet producing bias magnetic field of 760 Oe.

Table I. Summary of experimental data for spin wave interference: first column – measurement number; second column – antennas used for spin wave excitation. The numbers in the baskets correspond to the fraction of input power; third column - output power, forth column – output phase.

Table II. Summary of experimental data for individually working antennas: first column – measurement number; second column – antenna used for spin wave excitation. The numbers in the baskets correspond to the fraction of input power; third column- output power, forth column – output phase.

Figure 3. Experimental and theoretical data showing the phase of the output signal for the 20 sequential measurements. The red markers and red curve show the experimental data. The black markers and the curve show the theoretical data for the ideal case with zero input phase/amplitude variations.

Figure 4. (a) Experimental data showing the power of the output signal. The blue curve and blue markers depict the output power $P_{int}$ for the interfering spin waves (i.e. data from Table I). The green curve and green markers depict the sum power $P_{sum}$ as would be produced by the independently working antennas (i.e. data from Table II). (b) The res markers and the curve show the normalized output $P_{int}/P_{sum}$ for the 20 sequential measurements. The black markers and the curve depict the theoretical output in case of zero input phase/amplitude variations.

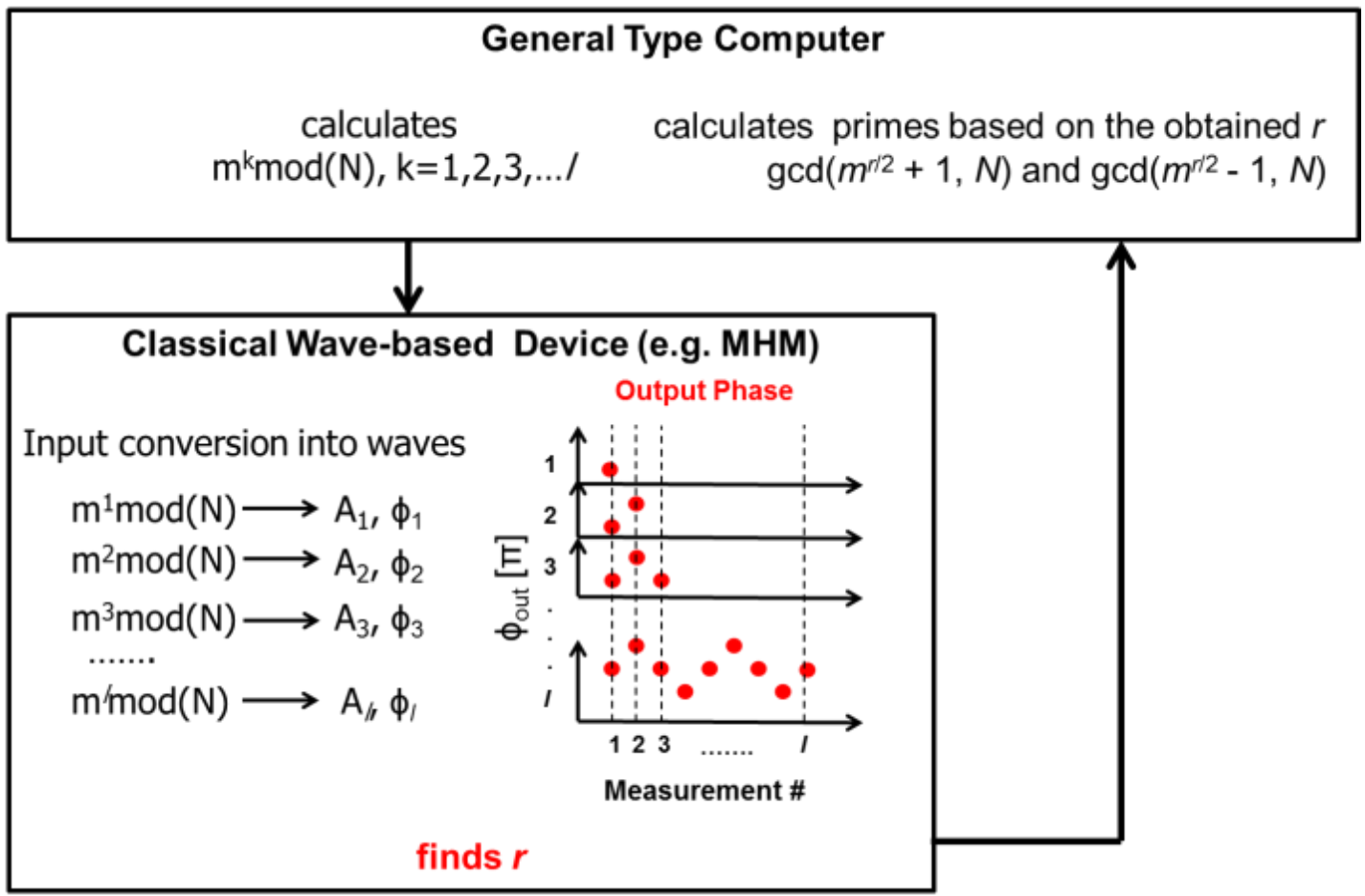

**Figure 1**

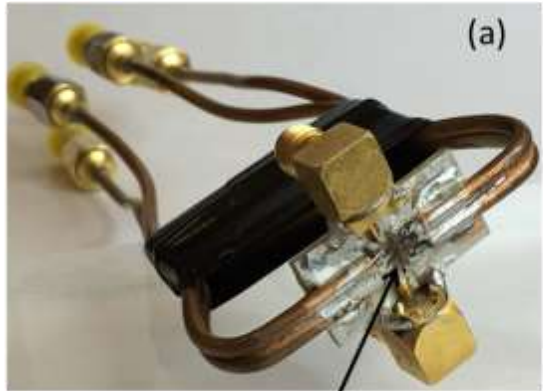
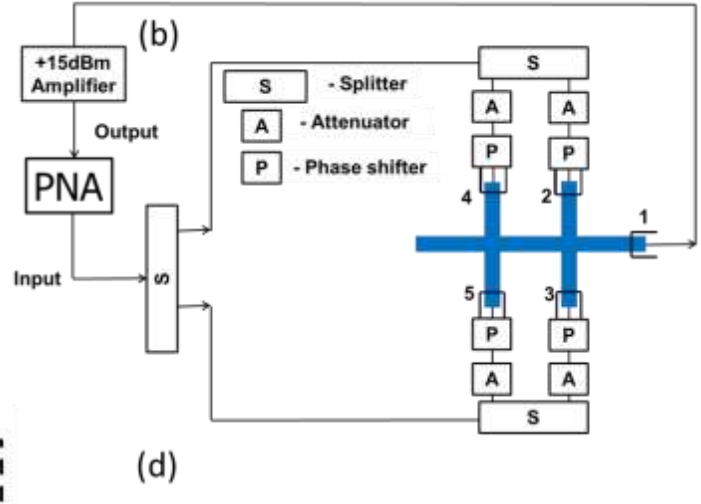
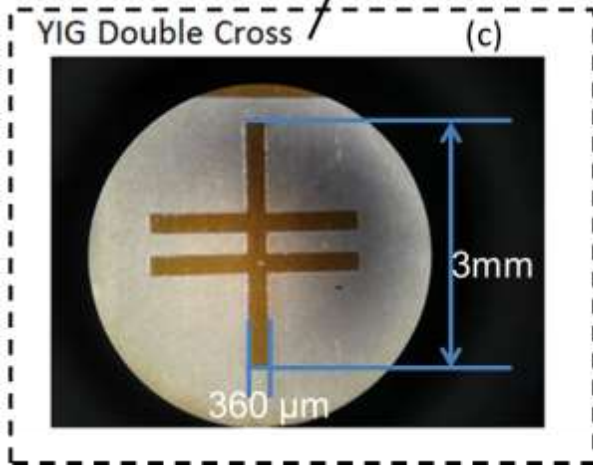
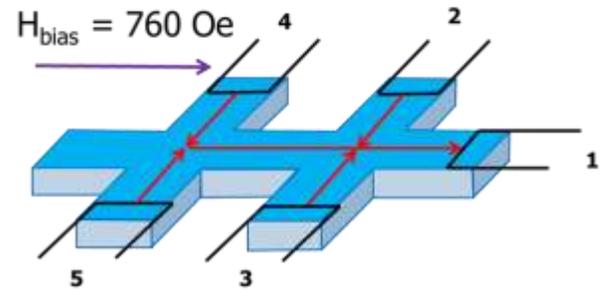

**Figure 2**

| Meas.# | Input antenna (amplitude change) | $S_{21}$ Power [mW] | $S_{21}$ Phase [degrees] |
|---|---|---|---|
| 1 | 2 | 0.581 | +2 |
| 2 | 2 + 4 | 1.07 | +11 |
| 3 | 2 + 4 + 3 | 1.56 | +25 |
| 4 | 2 + 4 + 3 + 5 | 1.88 | +42 |
| 5 | 2(1/2) + 4 + 3 + 5 | 1.72 | +45 |
| 6 | 2(1/2) + 4(1/2) + 3 + 5 | 1.50 | +55 |
| 7 | 2(1/2) + 4(1/2) + 3(1/2) + 5 | 1.23 | +50 |
| 8 | 2(1/2) + 4(1/2) + 3(1/2) + 5(1/2) | 1.04 | +45 |
| 9 | 2(1/4) + 4(1/2) + 3(1/2) + 5(1/2) | 0.961 | +50 |
| 10 | 2(1/4) + 4(1/4) + 3(1/2) + 5(1/2) | 0.850 | +55 |
| 11 | 2(1/4) + 4(1/4) + 3(1/4) + 5(1/2) | 0.679 | +54 |
| 12 | 2(1/4) + 4(1/4) + 3(1/4) + 5(1/4) | 0.585 | +45 |
| 13 | 2(1/8) + 4(1/4) + 3(1/4) + 5(1/4) | 0.547 | +50 |
| 14 | 2(1/8) + 4(1/8) + 3(1/4) + 5(1/4) | 0.482 | +55 |
| 15 | 2(1/8) + 4(1/8) + 3(1/8) + 5(1/4) | 0.390 | +53 |
| 16 | 2(1/8) + 4(1/8) + 3(1/8) + 5(1/8) | 0.338 | +44 |
| 17 | 2(1/16) + 4(1/8) + 3(1/8) + 5(1/8) | 0.313 | +50 |
| 18 | 2(1/16) + 4(1/16) + 3(1/8) + 5(1/8) | 0.274 | +53 |
| 19 | 2(1/16) + 4(1/16) + 3(1/16) + 5(1/8) | 0.221 | +56 |
| 20 | 2(1/16) + 4(1/16) + 3(1/16) + 5(1/16) | 0.194 | +47 |

**Table I**

| Meas. # | Input antenna (Amplitude change) | $S_{21}$ Power [mW] | $S_{21}$ Phase [degrees] |
|---|---|---|---|
| 1-4 | 2 | 0.581 | +2 |
| 2-5 | 4 | 0.576 | +30 |
| 3-6 | 3 | 0.609 | +59 |
| 4-8 | 5 | 0.589 | +90 |
| 5-9 | 2(1/2) | 0.305 | -2 |
| 6-10 | 4(1/2) | 0.325 | +30 |
| 7-11 | 3(1/2) | 0.363 | +61 |
| 8-12 | 5(1/2) | 0.327 | +90 |
| 9-13 | 2(1/4) | 0.165 | -3 |
| 10-14 | 4(1/4) | 0.186 | +28 |
| 11-15 | 3(1/4) | 0.202 | +60 |
| 12-16 | 5(1/4) | 0.185 | +90 |
| 13-17 | 2(1/8) | 0.91 | 0 |
| 14-18 | 4(1/8) | 0.106 | +30 |
| 15-19 | 3(1/8) | 0.117 | +58 |
| 16-20 | 5(1/8) | 0.104 | +87 |
| 17-20 | 2(1/16) | 0.51 | +1 |
| 18-20 | 4(1/16) | 0.62 | +28 |
| 19-20 | 3(1/16) | 0.66 | +61 |
| 20 | 5(1/16) | 0.60 | +90 |

**Table II**

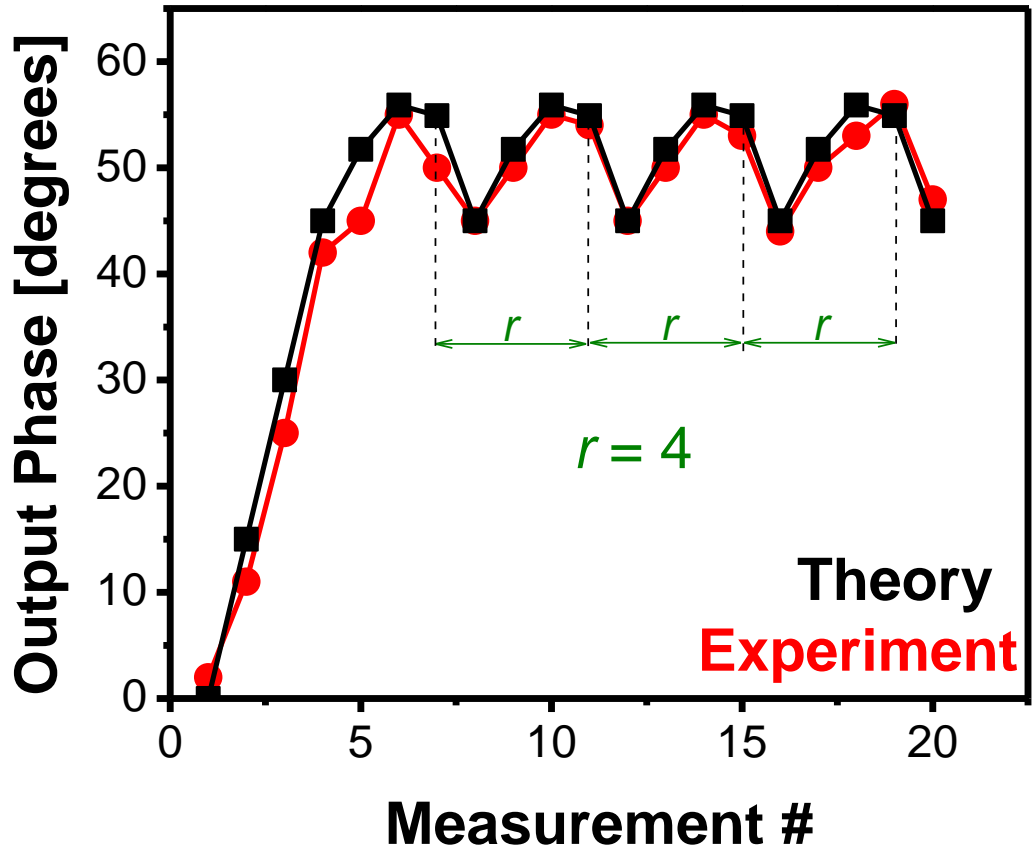

Figure 3

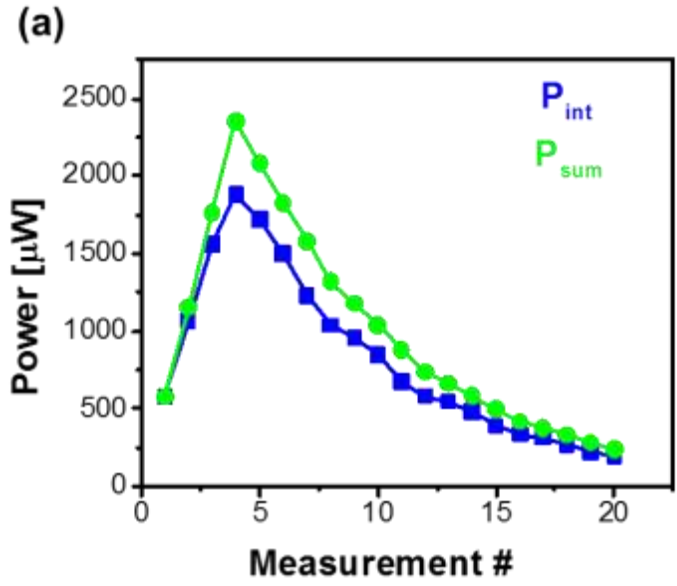 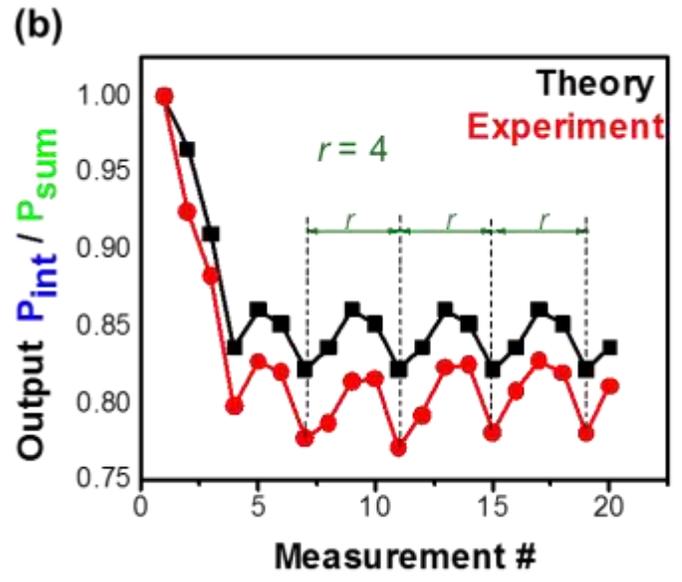

Figure 4